\documentstyle[12pt]{article}
\newcommand{\beq}{\begin{equation}}
\newcommand{\eeq}{\end{equation}}
\newcommand{\bea}{\begin{eqnarray}}
\newcommand{\eea}{\end{eqnarray}}
\newcommand{\bay}{\begin{array}}
\newcommand{\eay}{\end{array}}
\newcommand{\hc}{h_{v'}^{(c)}}                %charm-field
\newcommand{\hcbar}{\bar h_{v'}^{(c)}}        %charm bar-field
\newcommand{\hb}{h_{v}^{(b)}}                 %bottom-field
\newcommand{\hbbar}{\bar h_{v}^{(b)}}         %bottom bar-field

\newcommand{\vslash}{\mbox{$\not{\hspace{-1.03mm}v}$}}        % vslash
\newcommand{\dslash}{\mbox{$\not{\hspace{-1.03mm}D}$}}        % Dslash
\newcommand{\Dslash}{\mbox{$\not{\hspace{-1.03mm}D}$}}        % Dslash
\begin{document}
\begin{titlepage}
\begin{flushleft}
MZ-TH/93-13 (revised)\\
July 1993
\end{flushleft}
\begin{center}
\large
\bf
{\Large Heavy Quark Effective Theory at Large Orders in $1/m$}\\[2cm]
\rm
S.Balk$^1$,  J.G.K\"orner$^2$ and D.Pirjol$^1$\\[.5cm]
Johannes Gutenberg-Universit\"at\\
Institut f\"ur Physik (THEP), Staudingerweg 7\\
D-55099 Mainz, Germany\\[2cm]
\normalsize
\bf
Abstract\\
\rm
\small
\end{center}
   The existing derivations of a heavy quark effective theory (HQET) are
analyzed beyond the next-to-leading order in $1/m$. With one exception
they are found to be incorrect. The problem is a wrong
normalization of the heavy quark field in the effective theory. We argue
that the correct effective theory should be given by a Foldy--Wouthuysen
type field transformation to all orders in $1/m$. The renormalization of the
resulting Lagrangian to order $1/m^2$ is performed including also
effects arising through vacuum polarization. Our results for the
anomalous dimensions disagree with the existing ones. Some applications
are considered.
\\
[2cm]
\footnotesize
$^1\,$Supported by the Graduiertenkolleg Teilchenphysik, Universit\"at Mainz\\
$^2\,$Supported in part by the BMFT, FRG under contract 06MZ730\\
\normalsize
\end{titlepage}

\section*{1.Introduction}

  Interest in the physics of the mesons and baryons containing one heavy quark
($c$ or $b$) has received a strong stimulus through the invention of the
heavy quark effective theory (HQET) \cite{HQET}. This allows for a systematic
description of the symmetries appearing in the infinite quark mass limit
and of their breaking  by a finite quark mass.
An important outcome of this theory concerns the possibility
of accurately extracting the Kobayashi-Maskawa matrix element $V_{cb}$ in
a model-independent way through a study of the weak semileptonic decay
$\bar B\to D^*e\bar\nu$. This is made possible by  Luke's
theorem \cite{LUKE}, which asserts that some of the $1/m_c$--order corrections
to the infinite mass limit predictions of the theory for this decay vanish
when the $c$ quark is produced at rest in the $b$ quark's reference frame.
The next corrections come thus in at order $1/m_c^2$ and their estimation
is clearly important for an assessment of the reliability of the $V_{cb}$
determination. The task of evaluating these corrections has been recently
undertaken by Falk and Neubert \cite{FN}, who made use for this of the HQET
formalism developed in \cite{MRR,Lee}. The aim of this paper is to point
out that these treatments are incorrect, already at order $1/m^2$.
We show in Section 2 that the problem consists in a wrong normalization
of the field in the effective theory. The correct treatment turns out to
be the one given by the authors of \cite{KT}. However, due to the particular
strategy of calculation adopted in \cite{FN}, their results remain valid
when the correct treatment is being used.
In Sect.3 the matrix elements
of current operators obtained with the two HQETs are directly compared
and shown to differ at order $1/m^2$. A general presentation of the HQET
as given in \cite{KT} to any order in $1/m$ is offered in Sect.4.
The resulting effective Lagrangian to ${\cal O}(1/m^2)$ is used in Sect.5
to sum up
the leading logarithms of the heavy quark mass of the form $1/m^2\log(m)$
in the various matrix elements of the theory. We compute also the quantum
corrections appearing to this order from graphs containing one heavy
quark inside closed loops, the analog of the Euler--Heisenberg Lagrangian
relevant to the infrared behaviour of QCD.
An additional complication which appears when light quarks are included
is pointed out: light-quark operators are induced through renormalization.
In Sect.6 applications of the formalism to the phenomenology of heavy
hadrons are given.

\section*{2.Normalization of the field}

      Three distinct derivations of a heavy quark effective theory
(HQET) from QCD have been proposed  in the literature \cite{MRR,Lee,KT}.
Two of them \cite{MRR,Lee}, although phrased in different formalisms,
are actually equivalent. The third one \cite{KT} gives different results
from the first two, starting at order ${\cal O}(1/m^2)$. The aims of this
section are to explain the reasons for this disagreement and to show that
the last theory is the correct one.

   We start by first reviewing the results of the usual derivation
of the HQET as given in \cite{MRR}. The relation between the heavy quark
field in QCD, $Q(x)$, and that in the effective theory, $h(x)$, is
up to ${\cal O}(1/m^2)$
\beq\label{1}
 Q(x) = e^{-imv\cdot x}\left( 1 +\frac{1}{2m}(i\dslash_{\perp}) +
\frac{1}{4m^2}v\cdot D\dslash_{\perp}\right) h(x)\,.
\eeq
Here $v_{\mu}$ is the velocity of the heavy quark, $D_{\mu} =
\partial_{\mu}+igA_{\mu}^a t^a$ and $D_{\perp}^{\mu}=D^{\mu} -
v^{\mu}v\cdot D$. The heavy quark field $h(x)$ satisfies $\vslash h = h$.
The flavour-conserving vector current for quarks with
the same velocity can be written by making use of (1) as
\bea\label{2}
 \bar Q\gamma_{\mu}Q & = & \bar h\gamma_{\mu}h +\frac{i}{2m}\bar h
(\gamma_{\mu}\stackrel{\to}{\dslash_{\perp}} -
 \stackrel{\leftarrow}{\dslash_{\perp}}\gamma_{\mu})h \nonumber\\
&+&    \frac{1}{4m^2} \bar h
(\gamma_{\mu}v\cdot\stackrel{\to}{D}\stackrel{\to}{\dslash_{\perp}} +
\stackrel{\leftarrow}{\dslash_{\perp}}\gamma_{\mu}\stackrel{\to}
{\dslash_{\perp}} + \stackrel{\leftarrow}{\dslash_{\perp}} v\cdot
\stackrel{\leftarrow}{D} \gamma_{\mu})h\,.
\eea
The effective theory Lagrangian is
\bea\label{3}
{\cal L} = \bar h\left[ iv\cdot \stackrel{\to}{D} - \frac{1}{2m}
\stackrel{\to}{\dslash_{\perp}^2} + \frac{i}{4m^2}\stackrel{\to}
{\dslash_{\perp}}v\cdot \stackrel{\to}{D}\stackrel{\to}{\dslash_{\perp}}
\right] h\,.
\eea
Let us consider the number operator for heavy quarks at rest ($v_{\mu} =
(1,0,0,0)$, $\gamma_0 h(x) = h(x)$):
\bea\label{4}
\hat N = \int\mbox{d}^3\vec x\, (\bar Q\gamma_0 Q)(x) =
\int\mbox{d}^3\vec x\, \bar h(x)\left[ 1 + \frac{1}{4m^2}
\Dslash_{\perp}^2\right] h(x)\,.
\eea
Its expectation value in a hadronic state containing the heavy quark $Q$,
also at rest, is
\bea\label{5}
\lefteqn{\frac{1}{2m_M}\langle M_{QCD}(v)|\int\mbox{d}^3\vec x\,
(\bar Q\gamma_0 Q)(x)|M_{QCD}(v)\rangle =}\\
& &\frac{1}{2m_M}\langle M_{HQET}(v)|\int\mbox{d}^3\vec x \bar h_H(x)
\left[1 + \frac{1}{4m^2}\Dslash_{\perp}^2\right] h_H(x)|M_{HQET}(v)\rangle
\,.\nonumber
\eea
In this relation the field $h_H(x)$ is in the Heisenberg picture and its
evolution is dictated by the full HQET Lagrangian (3).
$|M_{QCD}(v)\rangle$ is an eigenstate of the QCD Lagrangian and
$|M_{HQET}(v)\rangle$ is the corresponding eigenstate of the full HQET
Lagrangian (3). Although they correspond to the same energy eigenvalue
(to the order in $1/m$ we are considering),
they are not necessarily identical and in fact, it will be shown in the
next section that they are related in the rest frame of the hadron by an
unitary transformation. The normalization convention for the states
$|M_{HQET}(v)\rangle$ is the usual one $\langle M_{HQET}(\vec p)|M_{HQET}
(\vec p\,')\rangle = 2E_{\vec p}\,\delta^{(3)}(\vec p - \vec p\,')$.

   The matrix elements on the r.h.s. of (5) can be written with the help
of the Gell-Mann--Low formula \cite{GML} as power series in $1/m$.
For example, the first term is equal to
\bea\label{6}
\langle M_{HQET}(v)|\int\mbox{d}^3\vec x (\bar h_H h_H)(x)
|M_{HQET}(v)\rangle = \langle M(v)|\int\mbox{d}^3\vec x(\bar h_I h_I)(x)
|M(v)\rangle \nonumber\\
+ \frac{1}{m}i \langle M(v)|\mbox{T}\int\mbox{d}^3\vec x
\int \mbox{d}^4y {\cal L_1}(y)(\bar h_I h_I)(x)|M(v)\rangle
+\cdots\,.
\eea
Here, the field $h_I(x)$ is understood to be in the interaction picture
where the term $\bar h(iv\cdot D)h$ in the Lagrangian (3) is the
unperturbed Lagrangian and the rest is an interaction. That is, we work to all
orders in $\alpha_s$; this is motivated by our desire to have hadrons
containing one heavy quark in the asymptotic initial and final states.
Another useful splitting of the Lagrangian (3) is the one which keeps only
the term $\bar hiv\cdot\partial h$ as the free Lagrangian and treats the rest
as a perturbation; this gives the usual double perturbative expansion in
$\alpha_s$ and $1/m$ and is suited for studying the scattering of free
quarks and gluons.
${\cal L}_1$ is the coefficient of $1/m$ in (3) and
the states $|M(v)\rangle$ are eigenstates of the leading order term
in the HQET Lagrangian.

    It can be seen that there are two distinct sources of mass dependence
in (5)\footnote{Neglecting the trivial $m$-dependence
appearing through the normalization of the states.}:
a)\,the factor $1/m^2$ appearing explicitly in (5), which has a kinematic
origin. It comes in through the very definition of a quantity of the
theory, the number operator. b)\,powers of $1/m$ arising through application
of the Gell-Mann--Low
formula. These can be considered as having a dynamical origin, as they
are due to the interaction terms in the Lagrangian (3).

   In the final expression obtained after application of the
Gell-Mann--Low formula to (5), the mass dependence is concentrated into
the factors of $1/m$ standing in front of matrix elements like those
on the r.h.s. of (6). The latter are mass-independent quantities,
properties of the HQET in the infinite mass limit. A similar expansion
as a power series in $1/m$ with constant coefficients can be written for
any other matrix element of heavy quark currents. In fact, it is this
separation of the mass dependence one of the properties which makes the
HQET useful.

   However, a closer look reveals some problems related to the form of
the heavy quark number operator (4). Firstly, its expectation value in
the unperturbed state $|M(v)\rangle$
\bea\label{7}
\frac{1}{2m_M}\langle M(v)|\int\mbox{d}^3\vec x\bar h_I(x)\left[ 1 +
\frac{1}{4m^2}\dslash_{\perp}^2\right] h_I(x)|M(v)\rangle
\eea
differs from unity. Usually, the scale of a quantized field is set by
requiring that the eigenvalues of the particle number operator be
integer numbers. In our case, the matrix element (7) can be normalized
to unity through the following field redefinition\footnote{A similar
transformation is sometimes made when deriving the Pauli equation to order
$1/m^2$. See e.g. \cite{Feynman}.}:
\beq\label{8}
h'(x) = \left[ 1-\frac{1}{2}\left( \frac{i\dslash_{\perp}}{2m}\right)^2
\right] h(x)\,.
\eeq

   Secondly, even more disturbing, the number operator (4) is not a
constant of motion for the leading order HQET Lagrangian:
\beq\label{9}
\frac{\mbox{d}}{\mbox{d}x_0}\int\mbox{d}^3\!x \bar h_I(x)\left[ 1 +
\frac{1}{4m^2}\dslash_{\perp}^2\right] h_I(x) \neq 0\,.
\eeq
To be sure, it {\em is} conserved when the dynamics is given by the full
Lagrangian (3). The point is that we do not know how to solve the
problem corresponding to this Lagrangian (nor can useful statements be
made about it). We are forced to
resort to the use of perturbation theory, which expresses any quantity of
interest in terms of matrix elements in the theory defined by the first
term in (3), the ``free'' theory. For these matrix elements, use can be
made of the symmetries of this theory in order to extract useful predictions.

Unfortunately, the nonconservation of the number operator (9) in the
``free'' theory has as a consequence the fact that the unperturbed states
$|M(v)\rangle$ upon which the perturbative expansion in $1/m$ is built,
simply do not exist. This means that the theory defined by (1--3) does
not lend itself to a consistent approximation scheme in powers of $1/m$.

Fortunately, the solution to the first problem, the field redefinition
(8), happens\footnote{That is, the same is not true at higher orders in
$1/m$.} to be also the solution for the second one. In terms of the
new field $h'(x)$, the Eqs.(1-3) read
\bea\label{10}
 Q(x) & = & e^{-imv\cdot x}\left[ 1 +\frac{1}{2m}(i\dslash_{\perp}) +
\frac{1}{4m^2}\left( v\cdot D\dslash_{\perp} - \frac{1}{2}
\dslash_{\perp}^2\right)\right] h'(x)\,,
\\
 \bar Q\gamma_{\mu}Q & = & \bar h'\gamma_{\mu} h' + \frac{i}{2m}
\bar h'(-\stackrel{\leftarrow}{\dslash_{\perp}}\gamma_{\mu}+\gamma_{\mu}
\stackrel{\to}{\dslash_{\perp}}) h'
\\ & & -
\frac{1}{4m^2}\bar h'\left( (\frac{1}{2}\stackrel{\leftarrow}
{\dslash_{\perp}^2} - \stackrel{\leftarrow}{\dslash_{\perp}}
v\cdot\stackrel{\leftarrow}{D})\gamma_{\mu} -
\stackrel{\leftarrow}{\dslash_{\perp}}\gamma_{\mu}
\stackrel{\to}{\dslash_{\perp}} + \gamma_{\mu}(\frac{1}{2}
\stackrel{\to}{\dslash_{\perp}^2} -
v\cdot\stackrel{\to}{D}\stackrel{\to}{\dslash_{\perp}})\right) h'\,,
       \nonumber\\
{\cal L'} &=& \bar h'\left[ iv\cdot \stackrel{\to}{D} - \frac{1}{2m}
\stackrel{\to}{\dslash_{\perp}^2}
 + \frac{i}{4m^2}\left(-\frac{1}{2}
\stackrel{\to}{\dslash_{\perp}^2}v\cdot \stackrel{\to}{D} +
\stackrel{\to}{\dslash_{\perp}}v\cdot\stackrel{\to}{D}\stackrel{\to}
{\dslash_{\perp}} - \frac{1}{2}v\cdot\stackrel{\to}{D}\stackrel{\to}
{\dslash_{\perp}^2}\right) \right] h'\,.\nonumber\\
\eea
This is precisely what one directly obtains by applying the method
described in \cite{KT} for deriving a heavy quark effective theory
from QCD. The number operator (5) is written now simply as
\beq\label{13}
\hat N = \int\mbox{d}^3\!x (\bar h' h')(x)
\eeq
and is a constant of motion for the free theory (i.e., when keeping only
the leading term in (12)).

   Equally important, it stays conserved up to each order in the
perturbative expansion in $1/m$ given by the Lagrangian (12). As discussed
above, this is a prerequisite for the consistency of the approximation
scheme. To see this to order $1/m$, we note that the equation of motion
for the field $h'(x)$ can be written as
\bea
\left[ iv\cdot \stackrel{\to}{D} - \frac{1}{2m}\stackrel{\to}
{\dslash_{\perp}^2}\right] h'(x)=0\,.
\eea
This can be used to write
\bea
\lefteqn{\frac{\mbox{d}}{\mbox{d}t}\hat N = \int\mbox{d}^3\vec x
   \left(\bar h'\stackrel{\leftarrow}{D}\cdot vh' +
   \bar h'v\cdot \stackrel{\to}{D}h'\right)(x) }\\
& &= \frac{i}{2m}\int\mbox{d}^3\vec x
   \left(\bar h'\stackrel{\leftarrow}{\dslash_{\perp}^2} h' -
   \bar h'\stackrel{\to}{\dslash_{\perp}^2} h'\right)(x) = 0\,,\nonumber
\eea
where an integration by parts has been performed, followed by the neglect
of the surface terms at infinity.

   The number operator for heavy quarks at rest is conserved also at
order $1/m^2$ in the effective Lagrangian (12). This can be seen most
easily by rewriting the $1/m^2$ term in the latter as
\bea
\frac{1}{2}\stackrel{\to}{\dslash_{\perp}}[v\cdot\stackrel{\to}{D} ,
\stackrel{\to}{\dslash_{\perp}}] + \frac{1}{2}[\stackrel{\to}
{\dslash_{\perp}}, v\cdot\stackrel{\to}{D}]\stackrel{\to}{\dslash_{\perp}}
=  \frac{ig}{2}\left( \stackrel{\to}{\dslash_{\perp}}v^{\mu}\gamma^{\nu}
F_{\mu\nu} - v^{\mu}\gamma^{\nu}F_{\mu\nu}\stackrel{\to}{\dslash_{\perp}}
\right)\,.
\eea
This gives an equation of motion for $h'(x)$
\bea
\left[ iv\cdot \stackrel{\to}{D} - \frac{1}{2m}\stackrel{\to}
{\dslash_{\perp}^2} - \frac{g}{8m^2}\left( \stackrel{\to}{\dslash_{\perp}}
v^{\mu}\gamma^{\nu}F_{\mu\nu} - v^{\mu}\gamma^{\nu}F_{\mu\nu}\stackrel{\to}
{\dslash_{\perp}}\right)\right] h'(x)=0\,,
\eea
which does not contain, in the rest frame of $v$, any time--derivative
acting on $h'(x)$, except in the leading--order term. The above procedure
can be repeated therefore, step by step, for this case as well.

   Everything which was said up to now refers to the number operator for
heavy quarks at rest. However, it can be seen that similar results
hold for the number operator for heavy quarks moving with an arbitrary
velocity $v$. Heavy quark effective theory (even truncated to a finite order
in $1/m$) is invariant under Lorentz transformations, provided the
velocity $v$ is transformed at the same time.
Furthermore the number operator is a Lorentz invariant and is the same in
any reference frame. In particular, if it is conserved in one reference
frame, it should also be conserved in any other reference frame.

\section*{3.Comparing matrix elements of the currents}

   We have shown in the preceding Section that the HQET (1-3) cannot
be formulated as a consistent perturbative expansion in powers of
$1/m$. In principle, this would suffice to reject it. However, we would
like to
present in this Section a different, more pragmatic point of view,
which will permit a better understanding of the difference between the
two HQETs, making at the same time the connection with the, perhaps
more familiar, theory of the point field transformations in the Lagrangian
formalism.

  We will compare in the following the explicit predictions of the two
HQETs, given
by (1-3) and respectively by (10-12), for the matrix element of the current
$\bar c\Gamma b$. Here $b,c$ are two different heavy quarks moving with
velocities $v,v'$ and having masses $m_b,m_c$.
By examining the form of the currents and Lagrangians, it is clear that a
possible difference can only show up at order ${\cal O}(1/m^2)$. This is
equal to
\bea
\lefteqn{\langle M_{HQET}'(v')|(\bar c\Gamma b)(0)|M_{HQET}(v)
\rangle_{(10-12)} -
\langle M_{HQET}'(v')|(\bar c\Gamma b)(0)|M_{HQET}(v)\rangle_{(1-3)}
 =}\nonumber\\
& &\frac{1}{8m_c^2}\langle M'(v')|\mbox{T}\int\mbox{d}^4x
  \hcbar(x)(\stackrel{\to}{\dslash_{\perp}'^2}v'\cdot\stackrel{\to}{D} +
  v'\cdot\stackrel{\to}{D}\stackrel{\to}{\dslash_{\perp}'^2})\hc(x)
(\hcbar\Gamma\hb)(0)|M(v)\rangle\nonumber\\
& & +
\frac{1}{8m_b^2}\langle M'(v')|\mbox{T}\int\mbox{d}^4x
  \hbbar(x)(\stackrel{\to}{\dslash_{\perp}^2}v\cdot\stackrel{\to}{D} +
  v\cdot\stackrel{\to}{D}\stackrel{\to}{\dslash_{\perp}^2})\hb(x)
(\hcbar\Gamma\hb)(0)|M(v)\rangle \\
& &-
\frac{1}{8m_c^2}\langle M'(v')|(\hcbar\stackrel{\leftarrow}
{\dslash_{\perp}'^2}\Gamma\hb)(0)|M(v)\rangle -
\frac{1}{8m_b^2}\langle M'(v')|(\hcbar\Gamma\stackrel{\to}
{\dslash_{\perp}^2}\hb)(0)|M(v)\rangle\,.\nonumber
\eea
We can rewrite this as
\bea
-\frac{i}{8m_c^2}\langle M'(v')|\mbox{T}\int\mbox{d}^4x
  \left( \hcbar\stackrel{\leftarrow}{\dslash_{\perp}'^2}
\frac{\mbox{d}{\cal L}}{\mbox{d}\hcbar} +
  \frac{\mbox{d}{\cal L}}{\mbox{d}\hc}\stackrel{\to}{\dslash_{\perp}'^2}
  \hc\right)(x) (\hcbar\Gamma\hb)(0)|M(v)\rangle\nonumber\\
-\frac{i}{8m_b^2}\langle M'(v')|\mbox{T}\int\mbox{d}^4x
  \left( \hbbar\stackrel{\leftarrow}{\dslash_{\perp}^2}
\frac{\mbox{d}{\cal L}}{\mbox{d}\hbbar} +
  \frac{\mbox{d}{\cal L}}{\mbox{d}\hb}\stackrel{\to}{\dslash_{\perp}^2}
  \hb\right)(x) (\hcbar\Gamma\hb)(0)|M(v)\rangle\nonumber\\
-\frac{1}{8m_c^2}\langle M'(v')|(\hcbar\stackrel{\leftarrow}
  {\dslash_{\perp}'^2}\Gamma\hb)(0)|M(v)\rangle -
  \frac{1}{8m_b^2}\langle M'(v')|(\hcbar\Gamma\stackrel{\to}
  {\dslash_{\perp}^2}\hb)(0)|M(v)\rangle\,,\nonumber\\
\eea
where ${\cal L}$ denotes the sum of the ``free'' Lagrangians of the two
heavy quarks:
\beq
{\cal L} = \hcbar iv'\cdot D\hc + \hbbar iv\cdot D\hb\,.
\eeq
The matrix elements (19) between hadronic states can be expressed with the
help of the reduction formalism as the residues at the corresponding
bound--state poles of the appropriate Green functions. For example, the
first term in (19) can be obtained from the Green function
\beq
\langle 0|\mbox{T}(\bar q\hc)(y)\int\mbox{d}^4x
  \left( \hcbar\stackrel{\leftarrow}{\dslash_{\perp}'^2}
\frac{\mbox{d}{\cal L}}{\mbox{d}\hcbar} +
  \frac{\mbox{d}{\cal L}}{\mbox{d}\hc}\stackrel{\to}{\dslash_{\perp}'^2}
  \hc\right)(x) (\hcbar\Gamma\hb)(0)(\hbbar q)(z)|0\rangle
\eeq
where $(\bar q\hc)$ and $(\hbbar q)$ are the interpolating fields of the
outgoing and respectively, incoming hadrons. Only their flavor structure
has been made explicit, the Lorentz and Dirac structure needed to endow
them with the correct quantum numbers of the respective bound states are
implicitly assumed.
One can apply now the relation\footnote{Given in \cite{Pol} and \cite{FGL}
with a wrong coefficient on the right--hand side. The correct form can be
found for example in \cite{Collins}.}
\bea
\lefteqn{\langle 0|\mbox{T}\left(F(\phi)\frac{\mbox{d}{\cal L}}
{\mbox{d}\phi}\right)(x)Q_1(x_1) Q_2(x_2)\dots |0\rangle =}\\
& & i\sum_i \delta^{(4)}(x-x_i)\langle 0|\mbox{T}Q_1(x_1)\dots
  \left(F\frac{\mbox{d}Q_i}{\mbox{d}\phi}\right)(x_i)\dots |0\rangle\,,
\nonumber
\eea
to transform (21) into
\bea
i\langle 0|\mbox{T}(\bar q\hc)(y)( \hcbar\stackrel{\leftarrow}
{\dslash_{\perp}'^2}\Gamma\hb)(0) (\hbbar q)(z)|0\rangle \\+
i\langle 0|\mbox{T}(\bar q\stackrel{\to}{\dslash_{\perp}'^2}\hc)(y)
( \hcbar\Gamma\hb)(0) (\hbbar q)(z)|0\rangle\,.\nonumber
\eea
The first term can easily be translated back into a matrix element
between hadronic states, equal to
\beq
i\langle M'(v')|( \hcbar\stackrel{\leftarrow}{\dslash_{\perp}'^2}
  \Gamma\hb)(0)|M(v)\rangle\,.
\eeq
In a completely analogous way, the second term in (19) can be transformed
to the form
\bea
i\langle 0|\mbox{T}(\bar q\hc)(y)
( \hcbar\Gamma\hb)(0) (\hbbar \stackrel{\leftarrow}{\dslash_{\perp}^2}
q)(z)|0\rangle \\+
i\langle 0|\mbox{T}(\bar q\hc)(y)( \hcbar\Gamma\stackrel{\to}
{\dslash_{\perp}^2}\hb)(0) (\hbbar q)(z)|0\rangle\,,\nonumber
\eea
of which the second term gives again a matrix element contributing in (18)
a quantity
\beq
i\langle M'(v')|( \hcbar\Gamma\stackrel{\to}{\dslash_{\perp}^2}
  \hb)(0)|M(v)\rangle\,.
\eeq
Collecting now everything, one can see that the terms (24) and (26) exactly
cancel the matrix elements of the two local terms in (18). The difference
(18) of the matrix elements of the current $\bar c\Gamma b$ in the two HQETs
is proportional to the residue at the corresponding bound--state poles of
the Green function
\bea
\frac{1}{8m_c^2}\langle 0|\mbox{T}(\bar q\stackrel{\to}
{\dslash_{\perp}'^2}\hc)(y)( \hcbar\Gamma\hb)(0) (\hbbar q)(z)|0\rangle\\
 + \frac{1}{8m_b^2}\langle 0|\mbox{T}(\bar q\hc)(y)
( \hcbar\Gamma\hb)(0) (\hbbar\stackrel{\leftarrow}{\dslash_{\perp}^2}
q)(z)|0\rangle\,.\nonumber
\eea
Next we observe that
\beq\label{28}
\dslash_{\perp}'^2\hc = \left( D^2-(v'\cdot D)^2+\frac{g}{2}\sigma_{\mu\nu}
F^{\mu\nu} + igv'_{\mu}\gamma_{\nu}F^{\mu\nu}\right) \hc\,.
\eeq
If we keep in this relation the terms containing {\em only} the heavy
quark field\footnote{The terms containing at least two fields do not
give a pole and vanish under multiplication with the LSZ factors.}, the
new interpolating field $(\bar q\hc)\to
(\bar q\partial_{\perp}^2\hc)$ has the same quantum numbers as
the old one and is therefore equally qualified to represent the same
hadronic bound states as the latter. A similar argument can be applied
to the second term in (27). We are led therefore to conclude that the
two Green functions in (27) have nonvanishing residues at the bound--state
poles associated with the hadronic states $|M(v)\rangle$ and
$|M'(v')\rangle$. Hence the two HQETs give different answers to order
${\cal O}(1/m^2)$.

  In taking the difference (18), the matrix elements on the r.h.s. are
taken between the same states. These are the physical states in the
infinite--mass limit. On physical grounds these states exist and
are unique. The reason for obtaining different results
with the two HQETs should be clear: in passing from one
theory to another, we have ``forgotten'' to make the change of
variable (8) in the interpolating field for the mesons too. Once
this is done, the contribution from the second term in (8) will
cancel the matrix elements (27). But this
means that the two interpolating fields $(\bar qh)$ and $(\bar
qh')$ cannot be simultaneously good interpolating fields for the
same states, to be used with their respective theories.
That is, we have an alternative: i) Use the HQET (1-3) with the
interpolating field $(\bar qh)$ for the heavy meson. Then, when working
with the HQET (10-12), the correct interpolating field to be used should be
\beq\label{29}
(\bar qh') - \frac{1}{8m^2}(\bar q\dslash_{\perp}^2h')\,.
\eeq
Or, ii) Use the HQET (10-12) with the interpolating field $(\bar qh')$
and the HQET (1-3) with the interpolating field
\beq\label{30}
(\bar qh) + \frac{1}{8m^2}(\bar q\dslash_{\perp}^2h)\,.
\eeq
Only one of these possibilities can be true (or none).
We will prove in the next section that the alternative ii) is the correct
one, by showing that the normalization of the interpolating field
$(\bar qh')$ is the same as that of its QCD counterpart $(\bar qQ)$:
\beq
 \langle 0 |(\bar qh'_H)|M_{HQET}\rangle =
 \langle 0 |(\bar qQ)|M_{QCD}\rangle \,.
\eeq

   The situation can be understood as a case of inapplicability of the
well--known equivalence
theorem \cite{EQUIV}. According to this theorem, any field transformation
$\phi' = \phi + F(\phi)$, where $F(\phi)$ contains {\em only terms with
at least two fields}, leaves the physical content of the theory
(the S--matrix) unchanged.
In other words, any arbitrary field $\phi'$ is as good as $\phi$ as
an interpolating field, as long as it has the same normalization:
\beq
 \langle 0 |\phi'(x)|\phi(p)\rangle =
 \langle 0 |\phi(x)|\phi(p)\rangle \,.
\eeq
The condition on the field transformation $F(\phi)$ has precisely the
function of assuring that the new field is correctly normalized.

 In our case, the two theories are related by the field transformation (8).
It is easy to see that the above-mentioned condition on the field
transformation is not satisfied for this case and therefore the two
fields $h(x)$ and $h'(x)$ do not have the same normalization. We will show
now that it is the field $h'(x)$ which is correctly normalized.

\section*{4.Proof to all orders in $1/m$}

  In this Section, after a short review of the results obtained in
\cite{KT}, we will take up the arguments of the preceding two Sections
and will prove that they are true to all orders in $1/m$.

  In the approach of \cite{KT}, the heavy quark field $h(x)$ in the
effective theory is expressed in terms of the QCD field $Q(x)$ as
\beq\label{16}
  h(x) = h^{(+)}(x) + h^{(-)}(x)
\eeq
with
\bea\label{17}
h^{(+)}(x) &=& \frac{1+\vslash}{2}e^{\textstyle imv\cdot x}\left(\cdots
e^{\textstyle -\frac{1}{2m^3}{\cal O}_3^A} e^{\textstyle -\frac{1}
{2m^2}{\cal O}_2^A}
e^{\textstyle -\frac{1}{2m}{\cal O}_1^A}\right) Q(x)\\
h^{(-)}(x) &=& \frac{1-\vslash}{2}e^{\textstyle -imv\cdot x}\left(\cdots
e^{\textstyle -\frac{1}{2m^3}{\cal O}_3^A} e^{\textstyle -\frac{1}
{2m^2}{\cal O}_2^A}
e^{\textstyle -\frac{1}{2m}{\cal O}_1^A}\right) Q(x)\,.
\eea
The differential operators ${\cal O}_i^A$ are determined so that the
Dirac Lagrangian
\beq\label{19}
{\cal L}_{Dirac} = \bar Q(i\Dslash - m)Q\,,
\eeq
does not contain, when expressed in terms of the fields $h^{(\pm)}(x)$,
any terms which couple $h^{(+)}(x)$ with $h^{(-)}(x)$. This is done by
removing order by order in $1/m$, through succesive field transformations,
every term in the Lagrangian which anticommutes with $\vslash$. These
represent the generalization of the ``odd'' operators in the usual
Foldy--Wouthuysen procedure. The first few
${\cal O}_i^A$ are
\bea\label{20}
{\cal O}_1^A &=& i\Dslash_{\perp}\\
{\cal O}_2^A &=& -\frac{1}{2}(\Dslash_{\perp}\Dslash_{\parallel} +
                            \Dslash_{\parallel}\Dslash_{\perp})\\
{\cal O}_3^A &=& -i(\frac{1}{3}\Dslash_{\perp}^3 +
  \frac{1}{4}\Dslash_{\perp}\Dslash_{\parallel}^2 +
  \frac{1}{2}\Dslash_{\parallel}\Dslash_{\perp}\Dslash_{\parallel} +
  \frac{1}{4}\Dslash_{\parallel}^2\Dslash_{\perp})\,.
\eea
All ${\cal O}_i^A$ satisfy $\{ {\cal O}_i^A , \vslash\} = 0$ and contain,
in the rest frame of $v$, only spatial derivatives which act on the $Q$
field in (34,35).

   It can be seen that this is a generalization of the usual
Foldy-Wouthuysen transformation which decouples the upper two components
of the Dirac spinor from the lower two ones. In a similar way, the
transformation (33-35) decouples those components of the Dirac spinor
which correspond to a particle moving with velocity $v$ from its
components corresponding to an antiparticle moving with the same velocity.
In the rest frame of $v$ the usual Foldy-Wouthuysen transformation is
recovered and the equation of motion for the Heisenberg field $h^{(+)}(x)$
becomes simply the Pauli equation. Motivated by the usual physical
interpretation
of the Foldy--Wouthuysen procedure, one can say in a loose way that
heavy quark effective theory represents a ``non-relativistic
approximation about some arbitrary velocity $v$''. This has been formulated
in a precise language to order $1/m^0$ by Grinstein \cite{Grinstein} (in
the rest frame of $v$).

   The recursive algorithm for determining the operators ${\cal O}_i^A$
and thereby the Lagrangian of the effective theory given in \cite{KT} is very
simple and lends itself easily to implementation in a symbolic
manipulation language (e.g. FORM \cite{FORM}). The effective Lagrangian
obtained in this way to order ${\cal O}(1/m^3)$ reads (only for the
particle field $h^{(+)}(x)$, which will be written simply as $h(x)$)
\beq\label{23}
{\cal L} = \sum_{i=0}^{\infty}\frac{1}{m^i}{\cal L}_i
\eeq
with
\bea\label{24}
{\cal L}_0 &=& \bar h iv\cdot Dh\\
{\cal L}_1 &=& -\frac{1}{2}\bar h\left( D^2 - (v\cdot D)^2 +
   \frac{g}{2}\sigma_{\mu\nu}F^{\mu\nu}\right) h\\
{\cal L}_2 &=& -\frac{g}{8}(\bar h v^{\mu}t^a h)D^{\nu}F^a_{\mu\nu}
   + \frac{ig}{8}(\bar h\sigma^{\alpha\nu}v^{\mu}t^a h)D_{\alpha}
   F^a_{\mu\nu} + \frac{ig}{4}(\bar h\sigma^{\alpha\nu}v^{\mu}F_{\mu\nu}
   D_{\alpha}h)\\
{\cal L}_3 &=& \frac{1}{8}\bar h \left( D^2-(v\cdot D)^2 +
   \frac{g}{2} \sigma^{\mu\nu}F_{\mu\nu}\right)^2 h\,.
% + g^2v_{\nu}v^{\alpha}[F^{\mu\nu},
%   F_{\mu\alpha}] \right.\\
%& &\left.\hoch{.6cm} - ig^2\sigma_{\mu\alpha}v_{\nu}v_{\beta}\{F^{\mu\nu},
%   F^{\alpha\beta}\}\right]h\,.\nonumber
\eea
The equation of motion for the $h(x)$ field in the rest frame of the
heavy quark is
\bea
\lefteqn{i\frac{\partial h}{\partial t} = (gA^{0a}t^a +
\frac{(\vec P - g\vec A^a
t^a)^2}{2m} - \frac{g}{2m}\vec\sigma\cdot\vec B^a t^a - \frac{g}{8m^2}
(\mbox{div}\vec E^a + gf_{abc}\vec A^b\cdot\vec E^c)t^a} \nonumber\\
& &-
\frac{ig}{8m^2}\vec\sigma\cdot\mbox{rot}\vec E^a t^a - \frac{ig^2}{8m^2}
f_{abc}\vec\sigma\cdot(\vec A^b\times\vec E^c)t^a - \frac{g}{4m^2}
\vec\sigma\cdot\vec E^a t^a \times (\vec P - g\vec A^b t^b)
\nonumber\\
& &-\frac{1}{8m^3}\left[ (\vec P - g\vec A^a t^a)^2 - g\vec\sigma\cdot
\vec B^a t^a \right]^2 + \frac{g^2}{8m^3}\left[ \vec E^a t^a\cdot
\vec E^bt^b + \frac{i}{2}f_{abc}\vec\sigma\cdot (\vec E^a\times\vec E^b)
t^c\right])h
\nonumber\\
& & +  {\cal O}(1/m^4)\,,
\eea
with $\vec P = -i\nabla$, $F_{\mu\nu}^a = \partial_{\mu}A_{\nu}^a -
\partial_{\nu}A_{\mu}^a - gf_{abc}A_{\mu}^bA_{\nu}^c$ and
$D_{\lambda}F_{\mu\nu}^a = \partial_{\lambda}F_{\mu\nu}^a - gf_{abc}
A_{\lambda}^bF_{\mu\nu}^c$. The electric and magnetic color fields are
defined as $E^{ia} = F^{i0a}$, $B^{ia} =
-\frac{1}{2}\epsilon^{ijk}F^a_{jk}$.
This agrees with the known form of the Pauli equation \cite{BjDr,Feinberg},
(though, to our knowledge, the complete ${\cal O}(1/m^3)$ contribution
is new) although the effort involved in deriving it, even in QCD, is
appreciably smaller than the one required by the traditional method using
unitary transformations. In the Appendix are given the operators
${\cal O}_{4,5}^A$ which allow the determination of the HQET Lagrangian
and of the Pauli equation up to ${\cal O}(1/m^5)$. The corresponding
expressions have been calculated up to ${\cal O}(1/m^{11})$ and will be
presented elsewhere \cite{AMON}.
\bigskip

  In Sect.~2 we have shown that the number operator in the HQET of
\cite{KT} has the simple form shown in Eq. (13). An important consequence
of this fact is that its expectation value in the unperturbed state
$|M(v)\rangle$ is equal to 1. Actually, this property is preserved to
all orders in $1/m$. To see this, we note from (33-35) that the
number operator can be written as
\bea
& &\hat N = \int\mbox{d}^3\vec x\,\bar Q\gamma_0 Q = \\
& & \int\mbox{d}^3\vec x\,\bar he^{\displaystyle imv\cdot x\vslash}\cdots
 e^{\displaystyle -\frac{1}{2m^2}\stackrel{\leftarrow}{{\cal O}_2^A}}
 e^{\displaystyle -\frac{1}{2m}\stackrel{\leftarrow}{{\cal O}_1^A}}
       \gamma_0
 e^{\displaystyle\frac{1}{2m}{\cal O}_1^A} e^{\displaystyle\frac{1}{2m^2}
 {\cal O}_2^A}\cdots
 e^{\displaystyle -imv\cdot x\vslash}h \nonumber\\
& & = \int\mbox{d}^3\vec x\,\bar he^{\displaystyle imv\cdot x\vslash}
   \gamma_0 e^{\displaystyle -imv\cdot x\vslash}h
= \int\mbox{d}^3\vec x\,(\bar h^{(+)}h^{(+)} - \bar h^{(-)}h^{(-)})(x)\,.
\nonumber
\eea
Here we have made use of the above-mentioned property of the differential
operators ${\cal O}^A_i$ of containing only spatial derivatives in order
to integrate by parts and drop the surface term. In the second step the
relation $\{ {\cal O}_i^A , \vslash\} = 0$ has been repeatedly used
(remember $\vslash = \gamma_0$). This proves the desired result.

  Another crucial property of the HQET in \cite{KT} which was proven
at order $1/m^2$ in Sect.~2 is the fact that the number operator in the full
theory is a constant of motion order by order in $1/m$. By making use
of the preceding result concerning the form of the number operator, one can
see that this is indeed the case
to any order in $1/m$. The proof for this is similar to the one
given in Eq. (15), where the main ingredient was the absence of the
time derivatives in the HQET Lagrangian (in the rest frame of $v$). It has
been proved in \cite{KT}
that the method of constructing the effective Lagrangian does not introduce
generally any $v\cdot\partial$ derivatives on the heavy quark field. Any
extra number
of derivatives can be reexpressed in terms of $F_{\mu\nu}$ and its
derivatives. Thus, the conservation law of $\hat N$ is guaranteed to hold
at any order in $1/m$.

  The conservation of the number operator order by order in $1/m$ has one
interesting consequence, in that its expectation value remains equal to
unity order by order in $1/m$. This follows from the nonrenormalization
theorem for conserved quantities, according to which matrix elements of
conserved operators have the same values as in the noninteracting theory.
We have just proved that the expectation value in the ``free'' theory
is unity, so that this is also the value to each order in $1/m$. This is,
of course, what one expects from the complete theory (QCD), and is
reassuring for the consistency of our scheme.

   The sequence of transformations (33-35) can be regarded as the
effect of an unitary transformation on the Hilbert space of the initial
(QCD) theory. Again, we consider the rest frame of $v$. The respective
unitary transformation is, of course, the usual Foldy--Wouthuysen
transformation in second--quantized form. We have
\beq
|M_{HQET}\rangle = U|M_{QCD}\rangle\,,\qquad\qquad
h = h^{(+)} + h^{(-)} = UQU^{\dagger}\,,
\eeq
with
\bea
U = \exp\left[-imx_0(N^{(+)}-N^{(-)})\right]
 \cdots \exp\left[ \frac{1}{2m^2}\int\mbox{d}^3x Q'^{\dagger}
  {\cal O}_2^A Q'\right]
\exp\left[ \frac{1}{2m}\int\mbox{d}^3xQ^{\dagger}
  {\cal O}_1^A Q\right]\,.
\eea
Here
\bea\label{31}
\lefteqn{Q'(\vec y) = e^{\displaystyle \frac{1}{2m}\int\mbox{d}^3x
 Q^{\dagger} {\cal O}_1^A Q} Q(\vec y) e^{\displaystyle -\frac{1}{2m}\int
 \mbox{d}^3xQ^{\dagger}{\cal O}_1^A Q} = e^{\displaystyle -\frac{1}{2m}
 {\cal O}_1^A}Q(\vec y)\,,}\\
\lefteqn{Q''(\vec y) = e^{\displaystyle\frac{1}{2m^2}\int\mbox{d}^3x
   Q'^{\dagger}{\cal O}_2^A Q'} Q'(\vec y) e^{\displaystyle -\frac{1}{2m^2}
   \int\mbox{d}^3xQ'^{\dagger}{\cal O}_2^A Q'} = e^{\displaystyle -\frac{1}
   {2m^2}{\cal O}_2^A}Q'(\vec y)\,,}\\
 & &\mbox{etc.}\,,\nonumber\\
\nonumber\\
\lefteqn{h^{(+)}(\vec y) + h^{(-)}(\vec y) =}\\
& &e^{\displaystyle -imx_0\int\mbox{d}^3x Q^{(\infty)\dagger}\vslash
   Q^{(\infty)}}Q^{(\infty)}(\vec y) e^{\displaystyle imx_0\int\mbox{d}^3x
   Q^{(\infty)\dagger}\vslash Q^{(\infty)}} = e^{\displaystyle imv\cdot x
   \vslash}Q^{(\infty)}(\vec y)\nonumber
\eea
give the effect of the successive transformations on the heavy quark field
operator and
\beq
N^{(\pm)} = \int\mbox{d}^3x\bar h^{(\pm)}h^{(\pm)}
\eeq
are the number operators for heavy particles and antiparticles, respectively.
In these relations all the operators are taken at equal times and $Q^{\dagger}
= \bar Q\gamma_0$. The second equality in Eqs.~(49-51) follows through
the use of the canonical anticommutation relations for the heavy quark
field. Here an important point is that these relations are left invariant
by the succesive unitary transformations.

  We can see now that the relation anticipated in the previous Section
\beq
\langle 0|(\bar qQ)|M_{QCD}\rangle =
\langle 0|(\bar qh)|M_{HQET}\rangle
\eeq
holds indeed true. This shows that the field $h(x)$ defined by the
sequence of transformations (49-51) has the same normalization as
the QCD field $Q(x)$ and therefore the results obtained with the HQET
formulated in terms of it are the correct ones. On the other hand, as
already remarked, the field corresponding to the usual HQET (1-3) has
a different normalization.

  One curious property of the effective Lagrangian (40) is the fact that
the number operators $N^{(+)}$ and $N^{(-)}$ for heavy quarks and
respectively, antiquarks, are separately conserved {\em to any order} in
$1/m$. This can be traced back to the invariance of the Lagrangian under the
field transformations $h\to \exp (i\alpha)h$ and $h\to \exp(i\vslash\alpha)
h$, the generators of which are related to the two number operators. But
this is surprising since we know that in the full theory (QCD), it is only the
difference of these two operators which is conserved (electric charge
conservation), but not each of them. That is, HQET has one more integral of
motion than the underlying theory. This seems to indicate the fact that
the predictions of HQET do not approach arbitrarily close those of QCD when
infinitely many orders in $1/m$ are taken into account. Namely, effects
connected with heavy quark pair production are not described, to all orders
in $1/m$.
See also Ref. \cite{Snyder}. We mention that this limitation applies
only to the tree-level HQET Lagrangian and it is possible that matching
corrections might remedy this deficiency. If not, this would signal a
limitation in the predictive power of HQET for very large orders in $1/m$.

\section*{5.Renormalization of the $1/m^2$ order Lagrangian}

      There is a class of contributions at order ${\cal O}(1/m^2)$
which are not described by the HQET Lagrangian (43). These arise in the
full theory
(QCD) from graphs containing one heavy quark in closed loops. They are
therefore quantum effects which have not been accounted for in the
tree level (classical) effective Lagrangian (40). Consequently, the
missing contributions must be added by hand as gluonic operators with
coefficients which have to be determined from a comparison (matching) with
the full QCD graphs. For simplicity we treat only the case when no light
quarks are present. There are three possible operators which can appear
when inserted at zero total momentum:
\bea\label{35}
{\cal O}_0 &=& \frac{1}{2}(D^{\mu}F_{\mu\nu}^a)(D_{\lambda}F^{a\lambda\nu})
 - \frac{1}{2} g(D^{\mu}F_{\mu\nu}^a)(\bar h\gamma^{\nu}t^a h)\\
{\cal O}_1 &=& \frac{1}{4} F_{\mu\nu}^a D^2 F^{a\mu\nu} \\
{\cal O}_2 &=& \frac{1}{4} gf_{abc} F^a_{\mu\nu}F^b_{\nu\rho}
   F^c_{\rho\mu}\,.
\eea
The operator ${\cal O}_0$ vanishes when the equations of motion for
the gluon field are used. By making use of the Bianchi identity
\beq\label{Bianchi}
 D_{\lambda}F_{\mu\nu}^a + D_{\mu}F_{\nu\lambda}^a +
 D_{\nu}F_{\lambda\mu}^a = 0\,
\eeq
one can see that the following relation between these operators
holds true, when inserted at zero momentum:
\beq\label{39}
  {\cal O}_1 + 2{\cal O}_2 +
  \frac{1}{2}(D^{\mu}F_{\mu\nu}^a)(D_{\lambda}F^{a\lambda\nu})= 0\,.
\eeq
The HQET Lagrangian (43) supplemented by these terms reads
\beq\label{40}
 {\cal L}_2 = \frac{1}{m^2}\sum_{i=1}^{4} c_i(\mu) {\cal O}_i\,.
\eeq
Here we have denoted by ${\cal O}_{3,4}$ the spin-symmetry violating
terms in (43):
\bea\label{41}
{\cal O}_3 &=& \frac{ig}{8}(\bar h\sigma^{\alpha\nu}v^{\mu}t^a h)D_{\alpha}
   F^a_{\mu\nu}\\
{\cal O}_4 &=& \frac{ig}{4}(\bar h\sigma^{\alpha\nu}v^{\mu}F_{\mu\nu}
   D_{\alpha}h)
\eea
Additional ${\cal O}(1/m^2)$ contributions also appear from double
insertions of ${\cal O}(1/m)$ terms in the effective Lagrangian:
\bea\label{43}
{\cal O}_5 &=& i\int\mbox{d}^4 x\,\mbox{T}(\bar h(iD)^2h)(x)(\frac{g}{2}
   \bar h\sigma\cdot F h)(0)\\
{\cal O}_6 &=& \frac{i}{2}\int\mbox{d}^4 x\,\mbox{T}(\frac{g}{2}
   \bar h\sigma\cdot F h)(x)(\frac{g}{2}\bar h\sigma\cdot F h)(0)\\
{\cal O}_7 &=& \frac{i}{2}\int\mbox{d}^4 x\,\mbox{T}(\bar h(iD)^2 h)(x)
   (\bar h(iD)^2 h)(0)
\eea
The coefficients $c_i(\mu)$ obey a renormalization group equation
\cite{EFF}
\beq\label{46}
 \mu\frac{\mbox{d}}{\mbox{d}\mu} c_i(\mu) + \gamma_{ji} c_j(\mu) = 0\,.
\eeq
The anomalous dimension matrix $\gamma_{ij}$ with $i,j=1-7$ has the form
\bea\label{47}
\hat\gamma = \frac{g^2}{16\pi^2} \left(\bay{ccccccc}
   -22 & -16  &  0  &  0  &  0  &  0  &  0 \\
    9  &   4  &  0  &  0  &  0  &  0  &  0 \\
    0  &   0  &-12  &  0  &  0  &  0  &  0 \\
    0  &   0  & 12  &  0  &  0  &  0  &  0 \\
    0  &   0  & 48  & 48  & -6  &  0  &  0 \\
    0  &   0  &  0  &  0  &  0  &-12  &  0 \\
    0  &   0  &  0  &  0  &  0  &  0  &  0 \\
 \eay\right) \,.
\eea
The nonlocal operators ${\cal O}_{5,6,7}$ are not required by the local ones
as counterterms and therefore their coefficients $c_{5,6,7}(\mu)$ evolve
independently from the latter. Their anomalous dimensions (the lower
$3\times 3$ diagonal block) have been calculated previously \cite{EH,FGL}.
The renormalization of the dimension--6 gluon operator ${\cal O}_2$ has been
studied in the presence of a light quark \cite{NT,M} in connection with the
renormalization of the corresponding condensate appearing in QCD sum rules.
We agree with the results of these works for the case of a massless quark.
However, when considering a HQET heavy quark, the renormalization is different.
The diagrams to be calculated for the renormalization of ${\cal O}_{1,2}$
are shown in Fig.1, those for ${\cal O}_{3,4}$ in Fig.2 and in Fig.3.
The necessary algebraic manipulations have been done with the help of FORM
\cite{FORM}. We have used for the
gluon field a Feynman version of the background field gauge \cite{Abbott},
where the combination $gA_{\mu}$ is not renormalized. In general, some
other operators will be needed as counterterms, which vanish when the
equation of motion for the heavy quark is used:
\bea\label{48}
{\cal O}_8 &=& \frac{ig}{2}\bar h\sigma\cdot F (v\cdot\stackrel{\to}{D})
  h - \frac{ig}{2} \bar h (\stackrel{\leftarrow}{D}\cdot v)\sigma\cdot Fh\\
{\cal O}_9 &=& i\bar h D^2 (v\cdot\stackrel{\to}{D})h
  - i\bar h (\stackrel{\leftarrow}{D}\cdot v)D^2h\\
{\cal O}_{10} &=& \bar h (iv\cdot D)^3 h\,.
\eea
Their contributions have to be properly separated. Even if these operators
do not contribute to the hyperfine splittings, they do appear when
considering corrections to the matrix elements of currents. Therefore, a
complete study of the ${\cal O}(\alpha_s/m^2)$ order corrections should
take them into account.

   The problem of the renormalization of the ${\cal O}(1/m^2)$ effective
Lagrangian has been treated previously by Lee \cite{Lee}. Our results
do not agree with his. For example, we find that the operators
${\cal O}_{3,4}$ are renormalized always in the combination
${\cal O}_3 + {\cal O}_4$, whereas in \cite{Lee} they appear together as
$2{\cal O}_3 + {\cal O}_4$ (denoted as $\hat O_2$). Unfortunately, this
combination is not even hermitian (whereas ${\cal O}_3 + {\cal O}_4$ is).
Its expectation values on the other hand have a direct physical
interpretation in terms of hyperfine level splittings which are clearly
real quantities.

   The coefficients $c_i$ at the matching scale $\mu = m$ are
\bea\label{51}
c_0(m) &=& -\frac{1}{4},\qquad c_1(m) = -\frac{1}{4} + \frac{g^2}{120\pi^2}
\nonumber\\
c_2(m) &=& -\frac{1}{2} + \frac{13g^2}{720\pi^2}, \qquad c_3(m) = 1\\
c_4(m) &=& 1, \qquad c_5(m) = -\frac{1}{4},\qquad c_6(m) = \frac{1}{4}\,.
\nonumber
\eea
Here the leading terms of the coefficients $c_{0,1,2}(m)$ have been
determined by reexpressing the first term in (43) in terms of
${\cal O}_{0,1,2}$ by making
use of (58). The contributions proportional to $g^2$ in $c_{1,2}(m)$
constitute the quantum corrections referred to at the beginning of this
section. They have been obtained by computing the QCD graphs in Fig.4,
where the fermion in the loop is a dynamical one of mass $m$.

   To one-loop order, the combinations of ${\cal O}_{1,2}$ which are
multiplicatively renormalized are
\bea
{\cal O}'_1 &=& {\cal O}_1 + 2{\cal O}_2\\
{\cal O}'_2 &=& 9{\cal O}_1 + 8{\cal O}_2
\eea
with the anomalous dimensions $\gamma'_1 = -\frac{g^2}{4\pi^2}\,,
\gamma'_2 = -\frac{7g^2}{8\pi^2}$ and the initial conditions
$c'_1(m) = -\frac{1}{4} + \frac{23g^2}{2400\pi^2}\,, c'_2(m) =
-\frac{g^2}{7200\pi^2}$. The matrix elements of ${\cal O}'_1$ can be
related via (58) to those of the first term in (43). Therefore,
as far as its matrix elements are concerned, the latter operator can
be considered as being multiplicatively renormalizable with the anomalous
dimension $\gamma'_1$. Summing up the leading logarithms
in $c'_{1,2}(\mu)$ and $c_{3,4}(\mu)$ with the help of the renormalization
group equation (65) and the initial conditions (70) gives the effective
Lagrangian ${\cal L}_2$ for an arbitrary $\mu < m$:
\bea
{\cal L}_2 &=& -\frac{\alpha_s(m)}{1800\pi}\left(\frac{\alpha_s(\mu)}
{\alpha_s(m)} \right)^{-\frac{7}{11}}{\cal O}'_2
+\left( \frac{\alpha_s(\mu)}{\alpha_s(m)} \right)^{-\frac{2}{11}}
\frac{g}{8}(\bar h v^{\nu}t^a h)D^{\mu}F^a_{\mu\nu}\nonumber\\
 &+&\left[2\left( \frac{\alpha_s(\mu)}{\alpha_s(m)}
\right)^{-\frac{3}{11}} - 1\right]({\cal O}_3 + {\cal O}_4)\,.
\eea
Here we have used the running of the coupling $\alpha_s$ proper to a
pure gauge field.

 It is interesting to observe that the first term is related to what we would
have obtained if we had ``completely'' integrated out the heavy quark. In the
absence of any other (light) quarks, there is only one independent dimension-6
operator which can appear into an Euler--Heisenberg Lagrangian for the gluon
field, which can be
chosen to be ${\cal O}_2$. Since now the gluon field is a free one,
Eq. (58) gives ${\cal O}'_2 = -10{\cal O}_2$ \footnote{This equality is to
be understood in the sense that the two operators have identical matrix
elements.}. Substituting this into (73) yields the well-known
result \cite{NSVZ}
\bea
{\cal L}^{E-H} = \frac{g^2}{720\pi^2 m^2}{\cal O}_2
+ {\cal O}(1/m^4)\,.
\eea
Also, $\gamma'_2$ is identical with the anomalous dimension of
${\cal O}_2$ for a pure gauge field \cite{NT}.

   However, keeping the first term in (73) is an inconsistent procedure,
as there are contributions of comparable magnitude into the
second term which were neglected. They are the non-leading logarithms in
$c'_1$ of the form $\alpha_s(\alpha_s\ln(\mu/m))^n$, whose summation by the
renormalization group will require the knowledge of the anomalous dimension
matrix (66) to two-loop order. This is somewhat a disappointing feature,
since, in the view of the preceding paragraph, they constitute the genuine
vacuum polarization contribution to the heavy quark effective Lagrangian.
Nevertheless, such a two-loop calculation is beyond the scope of this paper.

   Unfortunately, there are reasons to expect that the results for the
anomalous dimensions (66) will be
changed when light quarks are included. There are 19 new dimension--6
operators including both light- and heavy-quark fields and even purely
light-quark operators which can be induced by renormalization in the
presence of 3 massless quark species. Thus, there are 6 operators of
the form
\bea
& &(\bar q q)(\bar h h)\,,\quad (\bar q\gamma_{\mu} q)
(\bar hv_{\mu} h)\,, \quad (\bar q\sigma_{\mu\nu} q)
(\bar h\sigma_{\mu\nu} h)\nonumber\\
& &(\bar qt^aq)(\bar ht^a h)\,,\quad (\bar q\gamma_{\mu}t^aq)
(\bar hv_{\mu}t^a h)\,, \quad (\bar q\sigma_{\mu\nu}t^a q)
(\bar h\sigma_{\mu\nu}t^a h)
\eea
with the two light quark fields $\bar q q$, $q=(u\,,d\,,s)$ in a
SU(3)-singlet combination. There are 12 4-light-quark operators of
the following two types
\beq
 (\bar q\Gamma q)(\bar q\Gamma q)\,,\quad
 (\bar q\Gamma t^a q)(\bar q\Gamma t^aq)
\eeq
with two possibilities for each of them, corresponding to the two
occurences of {\bf 1} in the product of representations
({\bf 1}$\oplus${\bf 8})$\otimes$({\bf 1}$\oplus${\bf 8}). Here
$\Gamma = (1\,, \gamma_{\mu}\,, \sigma_{\mu\nu})$. And finally, 1
operator of the type $i\bar q \{F^{\mu\nu}, \gamma_{\mu}D_{\nu}\} q$.
For a similar calculation in QED with one light and one heavy fermion,
see \cite{QED}. Therefore the results of this section are mainly of
academic interest.

\section*{6.Applications}

        In this section we discuss some of the implications of
our results. One possible application would be to the hyperfine splittings
of hadrons containing one heavy quark. As is well-known, in the infinite
quark-mass limit $m_Q \to \infty$, the states of such hadrons can be
labelled by simultaneously specifying the angular momentum and parity
$s_{\ell}^{\pi_{\ell}}$ of its light degrees of freedom and the total
angular momentum $\vec S =\vec s_Q + \vec s_{\ell}$ (with $\vec s_Q$ the
heavy-quark spin). Since in this limit the spin of the heavy quark
decouples, there exists a degeneracy between the two states with the
same $s_{\ell}^{\pi_{\ell}}$ but different $S=s_{\ell}\pm\frac{1}{2}$
(for $s_{\ell} \neq 0$). The effect of a finite heavy quark mass is to
lift this degeneracy, leaving us with the observed hadronic spectrum
consisting of closely spaced doublets. The splittings within these
doublets are mainly a $1/m_Q$-effect due to the spin-symmetry violating
term in the effective Lagrangian ${\cal L}_1$ (42). At order $1/m_Q^2$,
an additional contribution is expected to appear from the corresponding
terms into ${\cal L}_2$ containing $\sigma_{\mu\nu}$. We can get some
insight into the size of these contributions by examining the role
played by these terms in the constituent quark model. In the rest frame
of the heavy-quark, they can be seen to be responsible for the
well-known spin-orbit interaction between the spin of the heavy quark
and its orbital angular momentum. Writing the spin-dependent part of
the Breit-Fermi Hamiltonian for a pair of a heavy- and light-quarks as
\cite{deRuGeGl}
\bea
 H = (\alpha e_Q e_q + k\alpha_s) (H^{spin-spin} + H^{spin-orbit})
\eea
with
\bea
\lefteqn{H^{spin-spin} = -\frac{8\pi}{3m_Q m_q}\delta^{(3)}(\vec r)
 \vec s_Q\cdot \vec s_q} \nonumber\\
 & &+ \frac{1}{2|\vec r|^3 m_Q m_q} \left(
\vec s_Q\cdot \vec s_q - 3\frac{(\vec s_Q\cdot r)(\vec s_q\cdot r)}
{|\vec r|^2} \right)\\
\lefteqn{H^{spin-orbit} = -\frac{1}{2|\vec r|^3 m_Q^2}\vec r\times \vec p_Q
\cdot \vec s_Q + \frac{1}{2|\vec r|^3 m_q^2}\vec r\times \vec p_q
\cdot \vec s_q} \nonumber\\
 & & -\frac{1}{|\vec r|^3 m_Q m_q}\vec r\times \vec p_Q
\cdot \vec s_q +\frac{1}{|\vec r|^3 m_Q m_q}\vec r\times \vec p_q
\cdot \vec s_Q\,,
\eea
then it is the first term in $H^{spin-orbit}$ which arises from the
above-mentioned terms in ${\cal L}_2$. Here $e_Q, m_Q, e_q, m_q$ are
respectively, the heavy and light quark electric charges and masses and
$k$ is $-\frac{4}{3}$ in mesons and $-\frac{2}{3}$ in baryons. The
constituent quark model suggests thus that the ${\cal L}_2$--contribution
to the mass splittings of s-wave mesons like $m_{D^*}-m_D$ or
$m_{B^*}-m_B$ is expected to be very small, being exactly zero in the
valence quark approximation\footnote{Of course, there are also $1/m^2$--order
contributions arising in second--order perturbation theory.
If one takes the usual point of view according to which the Breit-Fermi
Hamiltonian is only to be used as a first-order perturbation, they vanish
(for a review see \cite{Gromes}).}. This provides an explanation for the
fact that the observed hyperfine mass splittings of these mesons are
well described by a simple $1/m_Q$ law (including the leading order
logarithmic corrections) \cite{Lee}.
It should be manifest on the other hand in
the corresponding hyperfine splittings of the p-wave mesons which
therefore might display larger deviations from a simple $1/m_Q$ law.
Even for this case, the spin-spin forces will probably give the dominant
contribution.

  The ${\cal L}_2$-terms can be expected to play an important role for
the case of the first excited negative-parity
$I=0$ baryons with $s_{\ell}^{\pi_{\ell}}=1^-$. In a constituent quark
model
\cite{CIK,IWY} these states are described as having the light quark pair
($u,d$) into a spatially symetrical state and their total spin is zero.
If the inter-quark forces are taken to be of a harmonic oscillator type,
the relative angular momentum of the two light quarks is zero and the
resulting diquark ``looks'' like a scalar particle which orbits about
the heavy quark in a p-wave. When the heavy quark spin is added, two baryon
states will result $\Lambda^{\frac{1}{2}-}$ and $\Lambda^{\frac{3}{2}-}$.
The only term at ${\cal O}(1/m_Q)$ responsible for the hyperfine
splitting between these two states is the last one in $H^{spin-orbit}$
(79)\footnote{This statement is not specific to the harmonic force case.
It is only based on the symmetry of the wave-function under a permutation
of the two light quarks.}. One may thus expect the ${\cal L}_2$--term
(the first one in (79)) to contribute an appreciable fraction to the
hyperfine splitting of these baryons.

   Another application of our formalism is connected to the possibility
of a model-independent extraction of the Kobayashi-Maskawa matrix element
$V_{cb}$ as described in \cite{Neubert}. This method is based on the
absence of $1/m_c$ corrections to the form-factor $h_{A_1}(1)$ describing
the matrix element of the axial current $A_{\mu}=\bar c\gamma_{\mu}
\gamma_5 b$ for the transition $\bar B\to D^*e\bar\nu$:
\bea
\lefteqn{\langle D^*(v',\epsilon')|\bar c\gamma_{\mu}\gamma_5 b|
\bar B(v)\rangle =}\\
& & \sqrt{m_B m_{D^*}}\left[ h_{A_1}(v\cdot v')(1+v\cdot v')
\epsilon_{\mu}'^* - h_{A_2}(v\cdot v')(\epsilon'^*\cdot v)v_{\mu}
         - h_{A_3}(v\cdot v')(\epsilon'^*\cdot v)v'_{\mu}\right]\,.
\nonumber
\eea
In the infinite-mass limit, $h_{A_1}(1)=1$ and the corrections to this
result appear only at order $1/m_c^2$ \cite{LUKE}. It has been argued
in \cite{FN} that these corrections amount to about 2\%.
As mentioned in the Introduction, the calculation of \cite{FN} is based
upon the theory defined by (1) and (3). We have seen that these relations
are incorrect and have to be replaced by (10) and, respectively (12).
These changes imply a different form for the axial current in the HQET
which contains now two new contributions:
\bea
\Delta A_{\mu} = -\frac{1}{8m_c^2}\hcbar\stackrel{\leftarrow}
{\dslash_{\perp}}^2 \gamma_{\mu}\gamma_5\hb
                 -\frac{1}{8m_b^2}\hcbar\gamma_{\mu}\gamma_5
\stackrel{\to}{\dslash_{\perp}}^2 \hb\,,
\eea
proportional to $1/m_c^2$ and respectively $1/m_b^2$. It is natural to ask
how this change will affect the predictions of \cite{FN}. It turns out
that, due to the particular way of evaluating the $1/m^2$ corrections used
in \cite{FN}, they remain unchanged. A simple way to see this is presented
in the following. For notations we refer to \cite{FN}.

  The form--factor $h_{A_1}(1)$ at the equal-velocity point $v\cdot v' = 1$
is equal to
\beq
h_{A_1}(1) = 1 + \epsilon_c^2\ell_2(1) + \epsilon_b^2\ell_1(1) + \epsilon_c
   \epsilon_b(m_2(1) + m_9(1)) + {\cal O}(\epsilon^3)\,,
\eeq
with $\epsilon_Q = 1/(2m_Q)$ and $\ell_{1,2},\, m_{2,9}$ are functions
of $v\cdot v'$. Vector current conservation gives two conditions on the
values of these functions at the equal-velocity point $v\cdot v' = 1$:
\bea
& &2\ell_1(1) + (m_1(1) - m_8(1)) = 0\\
& &2\ell_2(1) + (m_4(1) - m_{11}(1)) + (m_5(1) - m_{12}(1)) = 0\,.
\eea
The ${\cal O}(1/m^2)$ correction to $h_{A_1}(1)$ is
\bea
h_{A_1}(1) - 1 = (\epsilon_c - \epsilon_b)[\epsilon_c\ell_2(1) -
   \epsilon_b\ell_1(1)] + \epsilon_c\epsilon_b\Delta
\eea
with
\beq
\Delta = \ell_1(1) + \ell_2(1) + m_2(1) + m_9(1)\,.
\eeq
The constants $\ell_{1,2}(1)$ are determined from a comparison with the
predictions of the Isgur, Scora, Grinstein and Wise model. This
determination is independent of the precise form of the HQET used.
$\Delta$ can be rewritten by making use of the relations (83-84) as
\beq
\Delta = -\frac{1}{2}(m_1(1) - m_8(1)) - \frac{1}{2}(m_4(1) - m_{11}(1)) -
          \frac{1}{2}(m_5(1) - m_{12}(1)) + (m_2(1)+m_9(1))\,.
\eeq

   Now it can be observed that the correction (81) only affects the
internal structure of the $\ell_{1-6}$ form--factors (see (4.34) in
\cite{FN}) and has no effect on the structure of the functions
$m_{1-24}$. Therefore the prediction for the value of the constant
$\Delta$ resulting from (87) remains valid.

  However, any calculation based on first principles of $\ell_{1-6}$,
such as with the help of QCD sum rules or on lattice, should give
an incorrect result when the HQET (1-3) is being used.

\section*{Acknowledgements}

We thank A.Ilakovac for collaboration in the early stages of this work
and to G.Karl, M.Lavelle and G.Thompson for a critical reading of the
manuscript. One of us (D.P.) is grateful to W.Kilian and T.Mannel for many
stimulating conversations and for hospitality during a visit to the
Technische Hochschule Darmstadt.

\section*{Appendix}

  We give here the differential operators ${\cal O}^A_{4,5}$, needed for
constructing the HQET Lagrangian up to ${\cal O}(1/m^5)$. They are
\bea
{\cal O}^A_4 &=& \frac{5}{16}\dslash_{\perp}^3\dslash_{\parallel}
   + \frac{3}{16}\dslash_{\perp}^2\dslash_{\parallel}\dslash_{\perp}
   + \frac{3}{16}\dslash_{\perp}\dslash_{\parallel}\dslash_{\perp}^2
   + \frac{1}{8}\dslash_{\perp}\dslash_{\parallel}^3\nonumber\\
  &+& \frac{5}{16}\dslash_{\parallel}\dslash_{\perp}^3
   + \frac{3}{8}\dslash_{\parallel}\dslash_{\perp}\dslash_{\parallel}^2
   + \frac{3}{8}\dslash_{\parallel}^2\dslash_{\perp}\dslash_{\parallel}
   + \frac{1}{8}\dslash_{\parallel}^3\dslash_{\perp}\\
{\cal O}^A_5 &=& i(\frac{1}{5}\dslash_{\perp}^5
   + \frac{7}{32}\dslash_{\perp}^3\dslash_{\parallel}^2
   + \frac{1}{4}\dslash_{\perp}^2\dslash_{\parallel}\dslash_{\perp}
     \dslash_{\parallel}\nonumber\\
  &+& \frac{3}{32}\dslash_{\perp}^2\dslash_{\parallel}^2\dslash_{\perp}
   + \frac{3}{16}\dslash_{\perp}\dslash_{\parallel}\dslash_{\perp}^2
     \dslash_{\parallel}
   + \frac{1}{8}\dslash_{\perp}\dslash_{\parallel}\dslash_{\perp}
     \dslash_{\parallel}\dslash_{\perp}
   + \frac{3}{32}\dslash_{\perp}\dslash_{\parallel}^2\dslash_{\perp}^2
\nonumber\\
  &+& \frac{1}{16}\dslash_{\perp}\dslash_{\parallel}^4
   + \frac{3}{8}\dslash_{\parallel}\dslash_{\perp}^3\dslash_{\parallel}
   + \frac{3}{16}\dslash_{\parallel}\dslash_{\perp}^2\dslash_{\parallel}
     \dslash_{\perp}
   + \frac{1}{4}\dslash_{\parallel}\dslash_{\perp}\dslash_{\parallel}
     \dslash_{\perp}^2\nonumber\\
  &+& \frac{1}{4}\dslash_{\parallel}\dslash_{\perp}\dslash_{\parallel}^3
   + \frac{7}{32}\dslash_{\parallel}^2\dslash_{\perp}^3
   + \frac{3}{8}\dslash_{\parallel}^2\dslash_{\perp}\dslash_{\parallel}^2
   + \frac{1}{4}\dslash_{\parallel}^3\dslash_{\perp}\dslash_{\parallel}
   + \frac{1}{16}\dslash_{\parallel}^4\dslash_{\perp})
\eea

\section*{Figure Captions}

{\bf Fig.1} Diagrams contributing to the anomalous dimension matrix of
${\cal O}_1$ and ${\cal O}_2$. (a) and (b) have a symmetry factor 1/2
and (f) appears together with another diagram obtained by reversing the
direction of the fermion line. In (d-g) the blob can only be an
insertion of ${\cal O}_1$.
\\[1cm]
{\bf Fig.2} Inserting ${\cal O}_3$ or ${\cal O}_4$ in heavy quark
vertex functions. Diagram (e) has a symmetry factor 1/2. Three other
diagrams contribute which are obtained from (a,b,d) by reversing the
direction of the fermion line.
\\[1cm]
{\bf Fig.3} Double insertions of ${\cal L}_1$. The square and the circle
may represent either different or identical terms in (42). To each diagram
which is asymmetrical with respect to its central vertical axis, its
mirror image should be added.
\\[1cm]
{\bf Fig.4} Diagrams needed for the matching conditions (70).

\end{document}